# Laser printing of silver and silver oxide

JORDAN M. ADAMS,[1*] DANIEL HELIGMAN[1], RYAN O-DELL[1], CHRISTINE Y. WANG[1], AND DANIEL YOUNG[2]

[1] *Riverside Research Institute, 2640 Hibiscus Way, Beavercreek, OH 45431, USA*

[2] *Wright State University,3640 Colonel Glenn Hwy, Dayton, OH 45435, USA*

[*] *jadams@riversideresearch.org*

**Abstract:** We show that direct laser writing in aqueous silver nitrate with a $\lambda$ = 1030 nm femtosecond laser results in deposition of a mixture of silver oxide and silver, in contrast to the pure silver deposition previously reported with 780 nm femtosecond direct laser writing. However, adding photoinitiator prevents silver oxide formation in a concentration-dependent manner. As a result, the resistivity of the material can also be controlled by photoinitiator concentration with resistivity being reduced from approximately 9e-3 Ωm to 3e-7 Ωm. Silver oxide peaks dominate the X-ray diffraction spectra when no photoinitiator is present, while the peaks disappear with photoinitiator concentrations above 0.05 wt%. A THz polarizer and metamaterial are printed as a demonstration of silver oxide printing.

## 1. Introduction

Multiphoton fabrication of metals is an additive manufacturing technique capable of printing high-reduction potential metals such as gold, platinum, palladium, or silver [1–11] . Typically, an 800 nm femtosecond laser is used to excite a photoinitiator with absorption around the two-photon wavelength of 400 nm, which then reduces dissolved metal ions [2–7]. Deposition may occur by the neutral atoms growing into nanoparticles that then aggregate onto the substrate [3,4].

While early work focused on the direct writing of metal from polymeric or sol-gel precursors [12, 13], more recent work has focused on fabrication in liquid environments. The first demonstration used an 800 nm femtosecond laser to print silver out of aqueous silver nitrate with no photoinitiator and achieved a resistivity value of 3.3 times that of the bulk silver value of 1.59e-8 Ωm [1]. Following works added photoinitiators [2] and surfactants [3] to the precursor to decrease the width of printed lines to <200 nm. A 1030 nm femtosecond laser has also been used to print silver from a photoinitiator/surfactant precursor and achieved similar widths [8].

While there is ample literature on the multiphoton printing of metals, the reported electrical conductance data and phase characterization is sparse. There are outstanding research questions regarding the phase of grown materials, especially when no photoinitiator is present and the deposition may be partially thermally induced [1,2]. Besides this, there are no reports of other silver compounds being printed, such as silver-oxide.

Here, we report on the fabrication of silver/silver-oxide using an aqueous $AgNO_3$ precursor solution containing varying amounts of photoinitiator precursor. Unlike the majority of reports which used an ~800 nm femtosecond laser, we used a 1030 nm femtosecond laser and 2-Benzyl-2-(dimethylamino)-4'-morpholinobutyrophenone (PI369) photoinitiator for 3-photon reduction. At 1030 nm with low photoinitiator (PI) concentrations, we show evidence that the printed structures are mixed metal/oxide with poor conductivity. Higher PI concentration results in pure metal deposition with the conductivity being four orders of magnitude higher. Additionally, energy dispersive spectroscopy (EDS) cross sections reveal the glass substrate is partially melted and mixed into the printed films only when no photoinitiator is present, further corroborating the transition from low-melting temperature silver oxide to silver. Observations reveal higher % PI tends to result in films with fewer holes and lines with less variability in width. Finally, we show two THz devices, a polarizer with variable line spacing that has an extinction ratio >750 and a split-ring resonator (SRR) array with fundamental resonance just below 0.4 THz.

This study offers insight into maximizing conductivity of laser printed silver by preventing silver oxide when unwanted. As this study solely focuses on understanding oxide growth, no surfactants were added to the precursor liquid. The lack of surfactants in combination with higher average power required for three-photon reduction limited the widths of lines to ~1 $\mu$m. However, this understanding of oxide growth should carry over to previously reported recipes which allow <200 nm widths. While adding additional photoinitiator or reducing agents may not further improve widths, there may be improvements in resistivity, especially for larger prints. Additionally, this study shows that printing silver oxide is possible, which can be valuable for applications involving supercapacitors [14], catalyzing specific reactions [15], as well as applications requiring specific bandgap properties [16].

## 2. Methods

An aqueous solutions of silver nitrate with a 1:1 mass ratio was prepared as the base precursor. Photoinitiator PI-369 (2-Benzyl-2-(dimethylamino)-1-[4-(4-morpholinyl)phenyl]-1-butanone) was added at concentrations between 0 wt% to 0.37 wt%, where the solution was fully saturated. A drop of the solution was placed on a borosilicate cover slip and a 1030 nm Yb-fiber femtosecond laser (~200 fs, 20 MHz) focused by 1.3 NA oil immersion objective was used to directly deposit material on the cover slip. After fabrication, the cover slip is rinsed in deionized water to remove residual photoinitiator and silver nitrate.

## 3. Experimental Results and Analysis

### *3.1. X-Ray Diffraction*

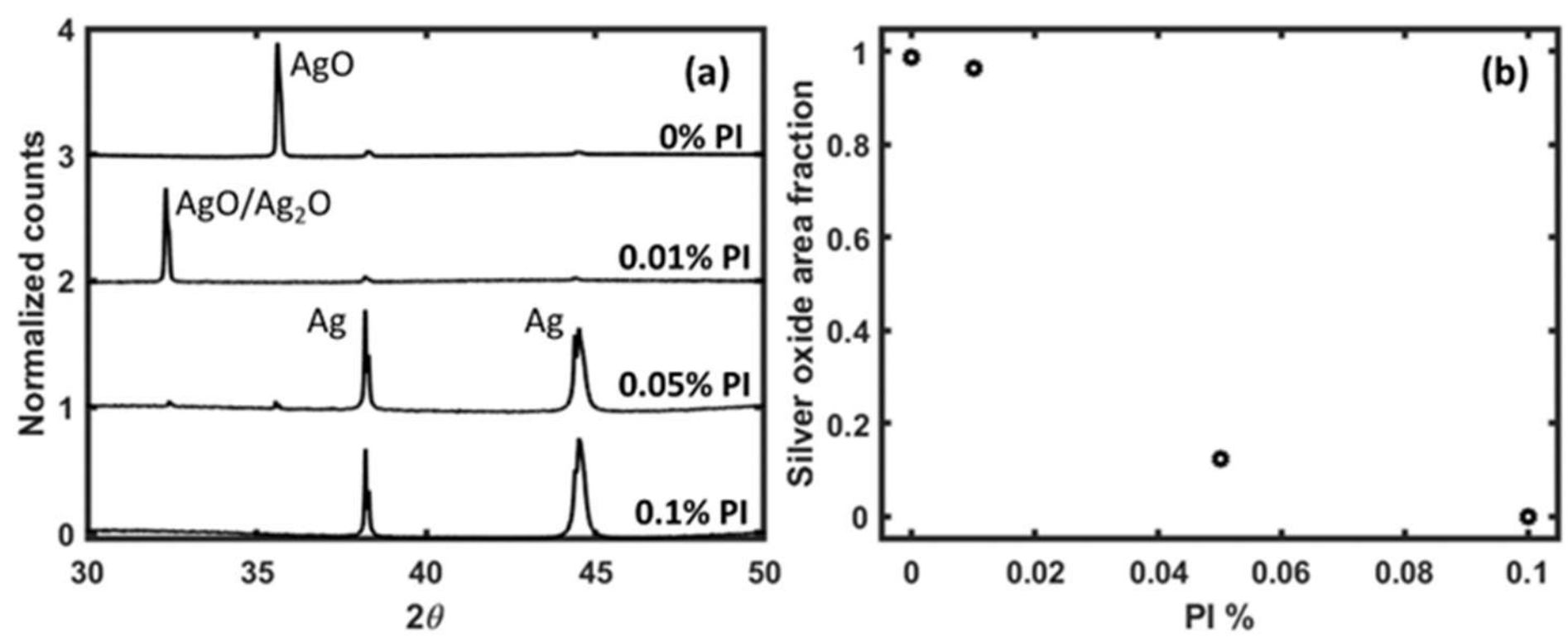

Fig. 1. (a) Normalized X-Ray Diffraction (XRD) measurements for each sample. (b) The fraction of area under the silver oxide peaks to the total area of the combined silver and silver oxide peaks.

Phase analysis was performed using a Rigaku Smartlab X-ray diffraction (XRD) system to determine the relative amounts of silver oxide and metallic silver. For these measurements, films were written in a raster pattern to provide sufficient material for analysis. The laser power was 600 mW with a scan velocity of 0.5 mm/s and raster steps of 1 $\mu$m. Figure 1 (a) shows the X-Ray Diffraction (XRD) results. The top line shows the XRD peak from the 0% photoinitiator (PI) sample is predominately [1 3 2] AgO (Ag(I)Ag(III)$O_2$) [17], with traces of metallic silver. At 0.01% PI the main peak becomes either AgO [1 2 1] or $Ag_2O$ [1 1 1], which have overlapping peaks. In both 0 and 0.01% PI samples, pure silver peaks are evident but still relatively small. At 0.05% PI concentration, the primary phase is pure silver, and the silver oxide peaks are relatively small. Finally, at 0.1% PI there are no noticeable silver oxide peaks remaining. The fraction of area under the two oxide peaks to the total area of the combined peaks is plotted in Figure 1 (b) for each concentration, which qualitatively shows amount of oxide being reduced with increasing concentration.

*3.2.* *Resistivity*

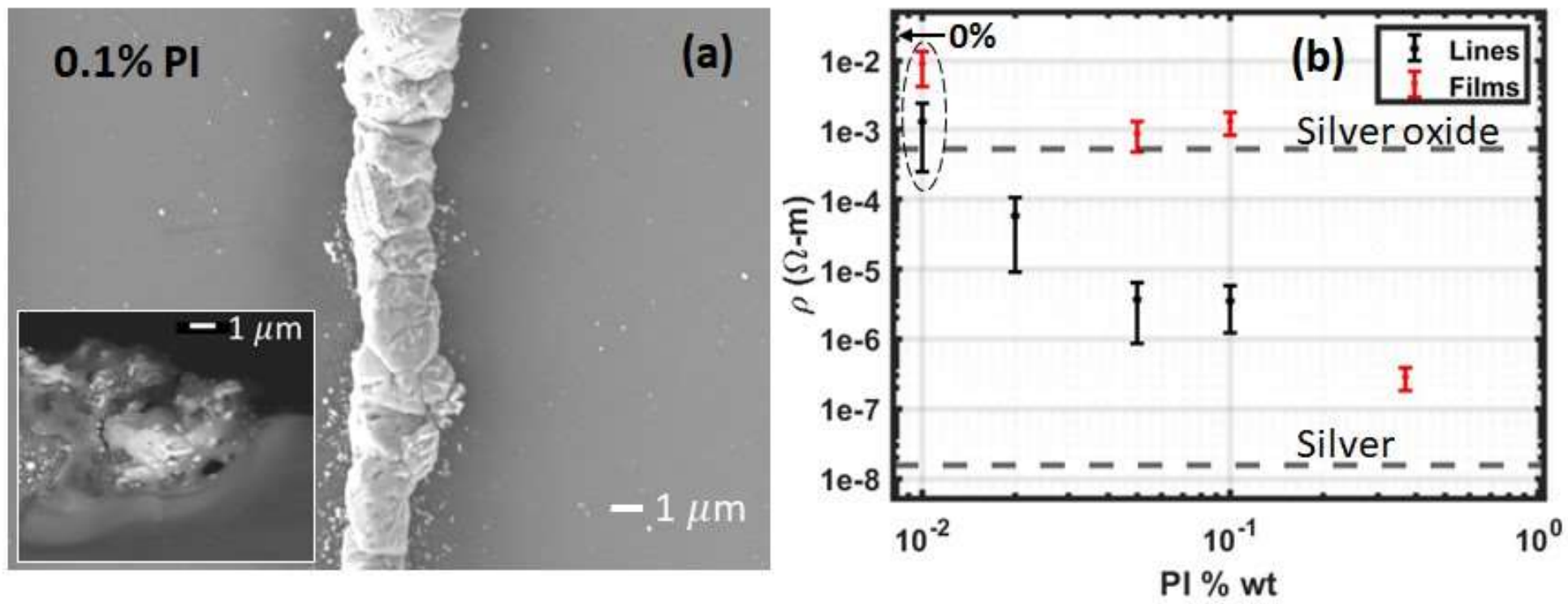

Fig. 2. (a) Example line written at 100 mW with 3 µm/s scan speed from a 0.1% PI solution and (b) calculated resistivity as a function of PI concentration for individual lines and films written in a raster pattern. The results for 0% PI are displayed on the log plot for comparison and highlighted with the dashed oval. The inset of Figure 2 (b) gives a side profile of a line written from a 0% PI solution, showing printed material is buried in the glass substrate.

Here, we present results showing the resistivity of the printed structures also decrease with increasing PI concentration. As porosity and variability in width decrease with increasing PI concentration, we note that this analysis gives an approximate effective resistivity and not intrinsic material resistivity. In order to study resistivity as a function of PI concentration, a Versalabs PPMS 4-point probe measurement system was used to measure the resistance. In addition to the films used for XRD analyses, single lines were printed between silver-paste electrode pads (Figure 2 (a)), utilizing a laser power of 100 mW power and 3 µm/s scan speed, where multiple passes were performed until a solid line became visible with optical microscope. The resistivity versus PI concentration for both lines and pads is shown in Figure 2 (b). SEM imaging (JEOL 7900 FLV) was also used to approximate line geometry for estimating the resistivity. There is variation in the width of the line, w, and height, h, across the length of the line, L, which complicate the resistivity calculation. We assume a circular cross section for lines and calculate the resistivity, $\rho = \frac{RA}{L} = R\frac{\pi w^2}{4L}$, for the maximum and minimum widths of lines, and plot the range of values with error bars on the plot. For films, the resistivity is $\rho = R\frac{wh}{L}$, where we use the respective height variation of the lines to plot a range of possible values with error bars. While previous reported work used a half-ellipse estimation for area of lines, which gives a lower resistivity [1], we found most lines and films are in fact partially buried in the glass (Figure 2 (a) inset) with lines not being a half-circle or ellipse lying flat on the substrate. As a reference, the resistivity of bulk silver oxide is 5.4e-4 Ωm [18] and bulk metallic silver is 1.58e-8 Ωm where both are plotted in Figure 2 as the thick upper and lower dashed lines, respectively. For the 0 wt% PI concentration, mean resistivity of the large film was 8.8e-3 Ωm, while the mean value for individual line was 1.3e-3 Ωm. With increasing PI concentration, the mean resistivity of the films drops to 2.8e-7 Ωm, ~18x that of bulk silver while the line resistivity drops to only 3.5e-6 Ωm. The decrease in resistivity with PI concentration is consistent with a shift from a silver oxide phase to a mixed metal/oxide, and finally to a pure metallic phase. PI 0.1 wt% is sufficient to achieve minimum line resistivity, while a significantly higher concentration of 0.37% PI achieved a similar resistivity value for the films. Section 3.3 will show that porosity is evident in the films printed at lower PI concentrations. Porosity may be a factor in why the films require a higher PI concentration to achieve minimal resistivity.

### *3.3. Observations from SEM, EDS, and Optical Images*

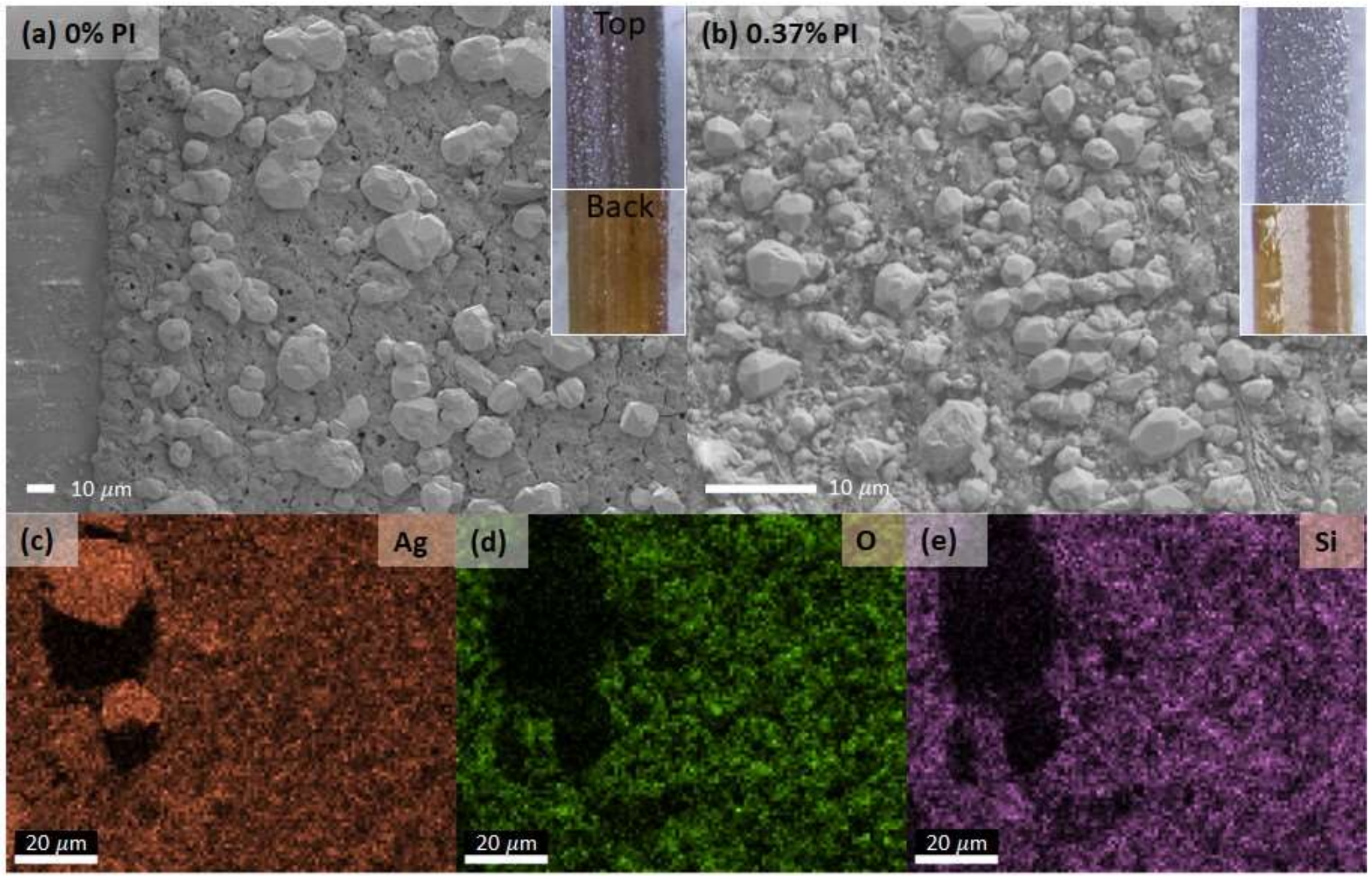

Fig. 3. SEM images of bulk films for (a) a 0 wt% PI sample and (b) a 0.37 wt% PI sample with insets showing optical images of the front and back side of the films. (c) Silver, (d) oxygen, and (e) silicon EDS maps of a small region of a 0% PI film with a crystallite.

Secondary-electron SEM images of the raster pattern films are shown in Figure 3, for 0% PI and 0.37% PI. In the 0% PI sample (Figure 3 (a)), the film has evidence of discrete pores and some cracking, while overlaid with crystallites. In the top inset, the optical image clearly shows the front side is brown colored which matches silver oxide, while some regions are also covered in highly reflective silver crystallites. The back side of the film, which is attached to the glass, is a mahogany color, which matches previous reports of boroscilicate doped with silver and may be evidence of silver diffusing into the substrate [19]. The 0.37% PI sample shows no porosity or cracking evident along with the overlaid crystallites. In this case, the optical image shows a grey film with matches silver, while the back side is the same mahogany color as the 0% PI sample. Figure 3 (c)-(e) show an EDS map of a crystallite and the film for a sample printed with no photoinitiator. While the crystallites are pure silver based on the EDS measurements, the films have varying concentrations of silicon and oxygen. The resistivity data is likely insensitive to the phase of the crystallites since they do not form a continuous electrical pathway. At the same time, we found EDS results for the 0.37% sample were similar, with silicon and oxygen appearing to be inside the film. However, due to the thinness of the film, the electron beam could be interacting with the substrate as well, giving rise to the measured oxygen and silicon.

To separate the substrate signal from the film signal, we performed EDS on cross sections show in Figure 4. Figure 4 (a)-(d) shows an SEM image of the cross-section of a 0% PI film as well as silver, oxygen, and silicon elemental maps. For clarity, a false blue color is given to the glass substrate region in the SEM image, while the interface is highlighted with a dashed line in the EDS images. Certain regions of the film have high concentrations of both silicon and oxygen at the same locations, while silver is well distributed. This indicates the glass may be partially mixing within the film. On the other hand, Figure 4 (e)-(h) shows the results for a 0.37% PI cross section. In this case, the oxygen and silicon are confined to the glass and not mixed with the film. The glass region of the SEM image is also given a false blue color, while the approximate location of the substrate faces are denoted with dashes lines in the EDS images. Since the film experienced no visible melting and mixing as before, this further corroborates the transition from low-melting temperature silver-oxide to pure silver with the addition of photoinitiator. Although not clearly shown, the 0.37% PI film is also partially buried below the original glass surface, while maintaining a distinct glass-silver interface.

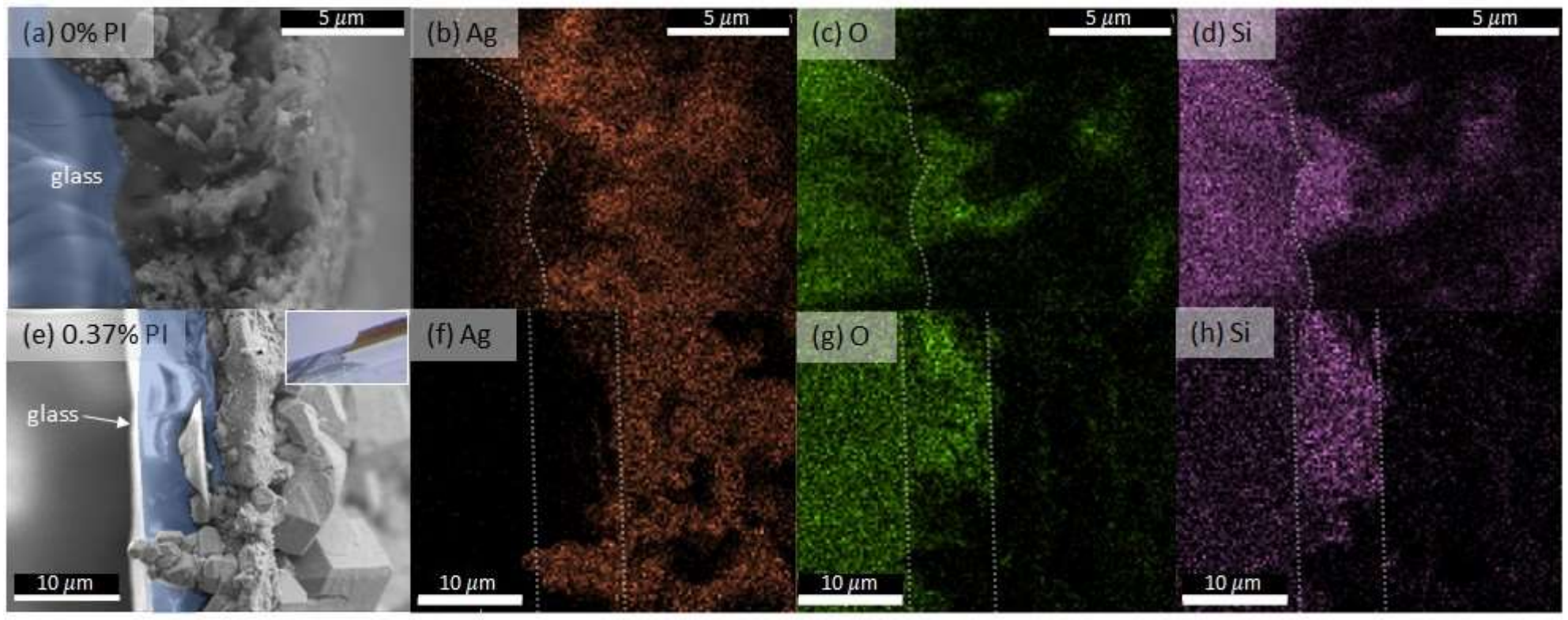

Fig. 4. SEM and EDS cross section images of films written from a (a)-(d) 0% PI solution and (e)-(h) 0.37% PI solution. The 0.37% PI cross section was taken from a region that naturally broke away from the cover slip and curled, as shown in the optical inset.

### *3.4. Discussion on Material Analysis and Potential Mechanisms of Oxide Growth*

Comparison of the EDS data, phase data, and resistivity data suggest that the material written from the 0% PI solution contains silver oxide while increasing the PI concentration reduces and ultimately eliminates oxide formation. When comparing these results of 1030 nm light photoinitiator-free printing to previously reported results using 800 nm, it is clear the center wavelength plays an important role in controlling oxide formation. Nonlinear interactions of the laser with the transiently forming metal nanoparticles, such as two-photon absorption, may offer one explanation for why silver and silver oxide deposition may depends on wavelength. Small silver nanoparticles formed via photoreduction would experience little plasmonic absorption and laser-induced heating with the 1030 nm femtosecond laser [20]. The low-temperature environment in the oxygen rich water could allow silver oxide growth on nanoparticles (NPs) as they slowly grow [21,22] and eventually deposit on the glass. In contrast, an 800 nm laser would more strongly excite silver nanoparticles owing to the second harmonic of the fundamental wavelength overlapping the 400 nm plasmonic peak [20], keeping them at a higher temperature and leading to more rapid growth. The hotter temperature of the particles and reduced time for growth could prevent oxide formation.

While the growing nanoparticles may be cooler with 1030 nm excitation, the deposited bulk material will reach higher temperatures. As a longer wavelength is used to excite higher-order multiphoton absorption, a higher laser power of at least 60 mW is needed at 1030 nm for deposition to occur compared to <15mW at 800 nm [1]. With the addition of the 3-photon PI, the deposition threshold drops down to 34 mW, which is still much higher average power than the two-photon based 800 nm deposition process. The higher laser power does mean that there will be significantly more single photon absorption once the structure is fabricated. The films will reach higher temperatures, and with no photoinitiator present, there will be local melting and mixing of the substrate into the low-melting temperature silver oxide. Additionally, the higher temperature of the films may lead to growth of the crystallites via thermal reduction. While femtosecond pulses are at first needed to initiate the deposition, we found continuous wave (CW) illumination could continue deposition of films when the laser beam profile was well overlapped with previously printed material. This occurred in our experiment of printing films with 1 $\mu$m raster steps or hatch distance. As an individual line is >1 $\mu$m in width, over half of the beam profile would overlap after a 1 $\mu$m or less raster step. In this case, a large portion of the beam energy can be absorbed by the previously printed material with single photon absorption. The films printed with CW light appeared to be visually similar in appearance to films printed using pulses for both with and without photoinitiator, although more testing is needed for confirmation. This suggests that a significant part of the printing process may actually be thermally driven at this laser power when there is modest overlap of the laser with the previously written material. In this case, the photoinitiator can also act as a general reducing agent that is thermally or chemically activated.

## 4. THz devices

To demonstrate silver oxide THz devices, we fabricated structures using the 0% PI aqueous silver nitrate solution. While the resistivity is several orders higher than that of pure silver, it is still low enough to have strong absorption and reflection in the THz band. A TeraMetrix THz Time-Domain Spectrometer was used to characterize the signal. For the first device, a polarizer was fabricated at different line spacings. The resulting perpendicular and parallel polarization reflections are shown in Figure 5 (a) and (b) with the locations of line spacings of 300 $\mu$m, 200 $\mu$m, 100 $\mu$m, 50 $\mu$m, and 25 $\mu$m are marked on an overlaid optical image. Figure 5 (c) shows an average cross-section for both polarizations and the corresponding extinction ratio which is the parallel divided by the perpendicular polarization reflection. While the perpendicular polarization is mostly transmitted with spacings of 50 $\mu$m and higher, the parallel component is significantly reflected. 50 $\mu$m spacing gave the best extinction ratio of >750 (or $\sim$29 dB), while the lines began to behave as a normal mirror at line spacings down to 25 $\mu$m. As a comparison, commercial THz wire-grid polarizers have extinction ratios of 30-50 dB [23], while research has shown >70 dB is possible for more complicated multi-layer wire-grid designs [24]. Due to the large size of the THz beam (0.3˜ mm), it is important to note the reflection from the edges of each region is partially overlapped with the adjacent section.

Next, we printed an electromagnetic split ring resonator (SRR) metamaterial array. Metamaterials have bulk optical properties dependent on the geometry of the unit cell [25]. The SRR metastructure can be viewed as an LC resonator in terms of an equivalent circuit model. The parametric length of the stripline, the dimensions of capacitive gap and the permeability/permittivity of ambient material dictate the structure's resonant frequency. Tuning this frequency will allow for changing the bulk metamaterial's optical properties [26,27]. Figure 6 (a) shows an SEM image of one part of the SRR array. The length of each side is 100 $\mu$m, for a total length of 300 $\mu$m. Since the substrate is 180 $\mu$m, there are Fabry-Perot resonances that overlap with the SRR resonances. Instead of showing the standard reflectivity, we divide the raw reflection spectrum by the reflection from the bare glass cover slip to eliminate the substrate interference resonances. Figure 6 (b) shows the results of the SRR array for both simulation and experiment. As the reported glass index in the THz regime varies strongly with the type of glass as well as impurities, we determined the index which gave the best alignment of the reflection peaks between simulation and experiment. This index value was n≈2.4 and the resulting simulation data is shown with the black line in Figure 6 (b). The index n≈2.4 was found to match the previously reported THz index of x=20 sodium borosilicate glass ($(x)Na_2O20B_2O_3$ $(80-x)SiO_2$) [28]. While the experimental peaks match well to simulation, the higher order peaks amplitude are significantly lower than simulation. The decrease in strength of the higher order resonances can be due to randomness in the surface profile at higher spatial frequency. This can arise from surface roughness, porosity, variation in the width of the lines, and randomly distributed particles over the glass surface. The addition of surfactants can be used to decrease the surface roughness while also preventing the growth of unwanted particles that may adhere as debris.

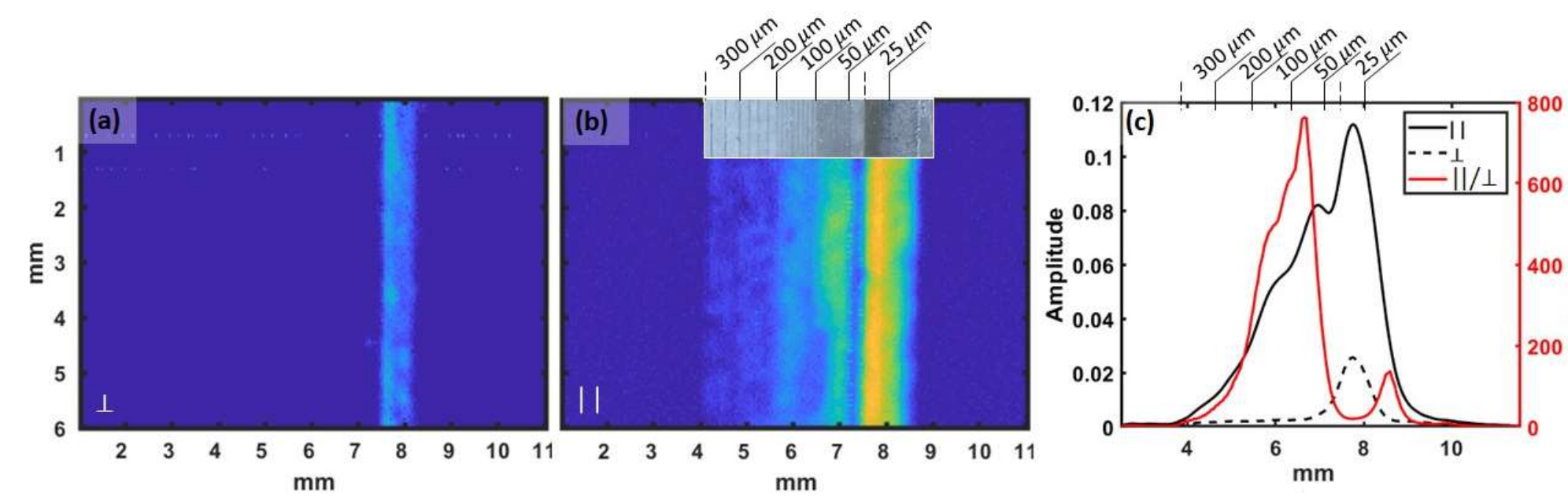

Fig. 5. (a), (b) Perpendicular and parallel polarization reflections from the polarizer. Inset on (b) is a optical image showing the black-brown lines at different spacings written on the coverslip. (c) Average cross-section reflection for both polarizations and the corresponding extinction ratio.

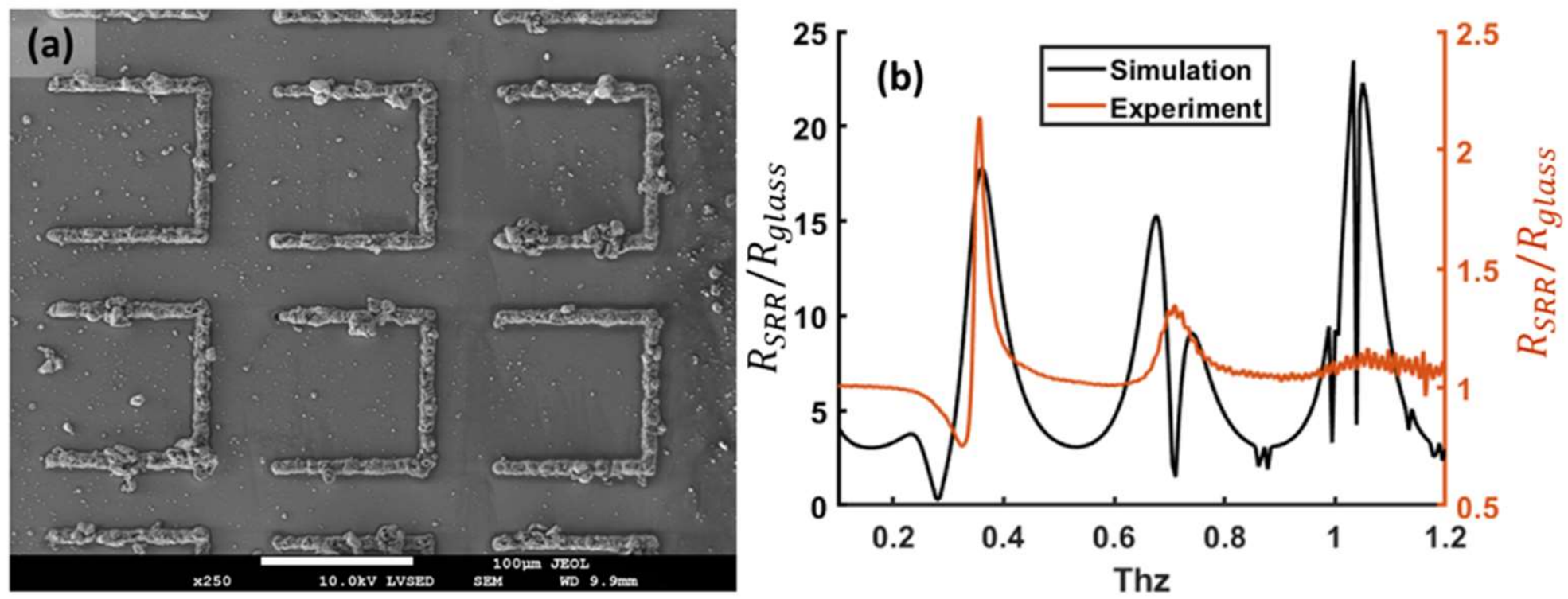

Fig. 6. (a) SEM image Split ring resonator (SRR) array unit cells. (b) Simulated and experimental reflection response from the SRR array.

## 5. Conclusion

In conclusion, we demonstrated that the amount of oxide in silver structures fabricated by direct laser writing from aqueous silver nitrate solutions can be controlled by the PI concentration when using a 1030 nm femtosecond laser. The resistivity of printed structures is controlled by photoinitiator concentration from approximately 9e-3 Ωm to 3e-7 Ωm. X-ray diffraction results show that silver oxide dominates the microstructure when no photoinitiator is present. Increasing concentrations of photoinitiator result in an increasing presence of metallic silver. A THz polarizer and metamaterial are shown as a demonstration of silver oxide printing. While further understanding of these growth mechanisms will be required to achieve the fabrication of optimized geometries, we show that careful control of PI concentration could allow precise fabrication of structures consisting of mixed metal/oxide materials.

Although using a 1030 nm femtosecond laser to print silver from a precursor, without a photoinitiator or with only a three-photon photoinitiator, requires higher average power and would result in significant single photon absorption, the precursor liquid is relatively transparent which should enable arbitrary geometry 3D printing. The understanding presented in this work should be directly applicable to improving resistivity of laser printing silver when using an 800 nm femtosecond laser or using a two-photon photoinitiator with 1030 nm light. In these cases, although increasing photoinitiator beyond the typical concentration may not further improve feature size, it could reduce resistivity, especially for larger prints. Future work should explore the effect of hatch distance on the contribution of single photon absorption and the resulting print quality. Additionally, a quantitative analysis of width variation, the number of defects, and surface roughness should be performed to determine the effect of variation in

% PI, scan speed and laser power. Determining whether scan speed and laser power effect the oxide content will also be valuable. Finally, while gold and platinum inherently do not form oxides, it would be valuable to determine if controllable oxide formation also occurs for lower reduction potential metals such as copper.

**Funding:** This work was supported with internal funding from Riverside Research Institute and Wright State University

**Disclosures:** The authors declare no conflicts of interest.

**Data Availability Statement:** The data that supports the finding of the current study are available from the corresponding author upon reasonable request.